\documentclass[amsmath,amssymb,aps,prl,twocolumn,superscriptaddress]{revtex4-1}

\usepackage{graphicx}
\usepackage{dcolumn}
\usepackage{bm}
\usepackage{xcolor}
\usepackage{subfigure}

\begin{document}

\title{Nuclear Mass Measurements Map the Structure of Atomic Nuclei \\
and Accreting Neutron Stars}

\author{Z.~Meisel}
\email[]{meisel@ohio.edu}
\affiliation{Institute of Nuclear \& Particle Physics, Department of Physics \& Astronomy, Ohio University, Athens, Ohio 45701, USA}
\author{S.~George}
\affiliation{Universit\"{a}t Greifswald, Institut f\"{u}r Physik, Greifswald 17487, Germany}
\author{S.~Ahn}
\affiliation{Cyclotron Institute, Texas A\&M University, College Station, Texas 77843, USA}
\author{D.~Bazin}
\affiliation{National Superconducting Cyclotron Laboratory, Michigan State University, East Lansing, Michigan 48824, USA}
\author{B.A.~Brown}
\affiliation{National Superconducting Cyclotron Laboratory, Michigan State University, East Lansing, Michigan 48824, USA}
\affiliation{Joint Institute for Nuclear Astrophysics -- Center for the Evolution of the Elements, Michigan State University, East Lansing, Michigan 48824, USA}
\affiliation{Department of Physics and Astronomy, Michigan State University, East Lansing, Michigan 48824, USA}
\author{J.~Browne}
\affiliation{National Superconducting Cyclotron Laboratory, Michigan State University, East Lansing, Michigan 48824, USA}
\affiliation{Joint Institute for Nuclear Astrophysics -- Center for the Evolution of the Elements, Michigan State University, East Lansing, Michigan 48824, USA}
\affiliation{Department of Physics and Astronomy, Michigan State University, East Lansing, Michigan 48824, USA}
\author{J.F.~Carpino}
\affiliation{Department of Physics, Western Michigan University, Kalamazoo, Michigan 49008, USA}
\author{H.~Chung}
\affiliation{Department of Physics, Western Michigan University, Kalamazoo, Michigan 49008, USA}
\author{R.H.~Cyburt}
\affiliation{National Superconducting Cyclotron Laboratory, Michigan State University, East Lansing, Michigan 48824, USA}
\affiliation{Joint Institute for Nuclear Astrophysics -- Center for the Evolution of the Elements, Michigan State University, East Lansing, Michigan 48824, USA}
\author{A.~Estrad\'{e}}
\affiliation{Department of Physics, Central Michigan University, Mount Pleasant, Michigan 48859, USA}
\author{M.~Famiano}
\affiliation{Department of Physics, Western Michigan University, Kalamazoo, Michigan 49008, USA}
\author{A.~Gade}
\affiliation{National Superconducting Cyclotron Laboratory, Michigan State University, East Lansing, Michigan 48824, USA}
\affiliation{Department of Physics and Astronomy, Michigan State University, East Lansing, Michigan 48824, USA}
\author{C.~Langer}
\affiliation{Department of Energy Technology, University of Applied Science Aachen, Campus J\"{u}lich, 52428 J\"{u}lich, Germany}
\author{M.~Mato\v{s}}
\affiliation{Physics Section, International Atomic Energy Agency, Vienna 1400, Austria}
\author{W.~Mittig}
\affiliation{National Superconducting Cyclotron Laboratory, Michigan State University, East Lansing, Michigan 48824, USA}
\affiliation{Department of Physics and Astronomy, Michigan State University, East Lansing, Michigan 48824, USA}
\author{F.~Montes}
\affiliation{National Superconducting Cyclotron Laboratory, Michigan State University, East Lansing, Michigan 48824, USA}
\affiliation{Joint Institute for Nuclear Astrophysics -- Center for the Evolution of the Elements, Michigan State University, East Lansing, Michigan 48824, USA}
\author{D.J.~Morrissey}
\affiliation{National Superconducting Cyclotron Laboratory, Michigan State University, East Lansing, Michigan 48824, USA}
\affiliation{Department of Chemistry, Michigan State University, East Lansing, Michigan 48824, USA}
\author{J.~Pereira}
\affiliation{National Superconducting Cyclotron Laboratory, Michigan State University, East Lansing, Michigan 48824, USA}
\affiliation{Joint Institute for Nuclear Astrophysics -- Center for the Evolution of the Elements, Michigan State University, East Lansing, Michigan 48824, USA}
\author{H.~Schatz}
\affiliation{National Superconducting Cyclotron Laboratory, Michigan State University, East Lansing, Michigan 48824, USA}
\affiliation{Joint Institute for Nuclear Astrophysics -- Center for the Evolution of the Elements, Michigan State University, East Lansing, Michigan 48824, USA}
\affiliation{Department of Physics and Astronomy, Michigan State University, East Lansing, Michigan 48824, USA}
\author{J.~Schatz}
\affiliation{National Superconducting Cyclotron Laboratory, Michigan State University, East Lansing, Michigan 48824, USA}
\author{M.~Scott}
\affiliation{National Superconducting Cyclotron Laboratory, Michigan State University, East Lansing, Michigan 48824, USA}
\affiliation{Department of Physics and Astronomy, Michigan State University, East Lansing, Michigan 48824, USA}
\author{D.~Shapira}
\affiliation{Oak Ridge National Laboratory, Oak Ridge, Tennessee 37831, USA}
\author{K.~Smith}
\affiliation{Los Alamos National Laboratory, Los Alamos, New Mexico 87545, USA}
\author{J.~Stevens}
\affiliation{National Superconducting Cyclotron Laboratory, Michigan
State University, East Lansing, Michigan 48824, USA}
\affiliation{Joint Institute for Nuclear Astrophysics -- Center for the Evolution of the Elements, Michigan State University, East Lansing, Michigan 48824, USA}
\affiliation{Department of Physics and Astronomy, Michigan State University, East Lansing, Michigan 48824, USA}
\author{W.~Tan}
\affiliation{Department of Physics, University of Notre Dame, Notre Dame, Indiana 46556, USA}
\author{O.~Tarasov}
\affiliation{National Superconducting Cyclotron Laboratory, Michigan State University, East Lansing, Michigan 48824, USA}
\author{S.~Towers}
\affiliation{Department of Physics, Western Michigan University, Kalamazoo, Michigan 49008, USA}
\author{K.~Wimmer}
\affiliation{Department of Physics, University of Tokyo, Hongo 7-3-1, Bunkyo-ku, Tokyo 113-0033, Japan}
\author{J.R.~Winkelbauer}
\affiliation{Los Alamos National Laboratory, Los Alamos, New Mexico 87545, USA}
\author{J.~Yurkon}
\affiliation{National Superconducting Cyclotron Laboratory, Michigan State University, East Lansing, Michigan 48824, USA}
\author{R.G.T.~Zegers}
\affiliation{National Superconducting Cyclotron Laboratory, Michigan State University, East Lansing, Michigan 48824, USA}
\affiliation{Joint Institute for Nuclear Astrophysics -- Center for the Evolution of the Elements, Michigan State University, East Lansing, Michigan 48824, USA}
\affiliation{Department of Physics and Astronomy, Michigan State University, East Lansing, Michigan 48824, USA}

\date{\today}

\begin{abstract}
    We present mass excesses (ME) of neutron-rich isotopes of Ar through Fe, obtained via TOF-$B\rho$ mass spectrometry at the National Superconducting Cyclotron Laboratory. Our new results have significantly reduced systematic uncertainties relative to a prior analysis, enabling the first determination of ME for $^{58,59}{\rm Ti}$, $^{62}{\rm V}$, $^{65}{\rm Cr}$, $^{67,68}{\rm Mn}$, and $^{69,70}{\rm Fe}$. Our results show the $N=34$ subshell weaken at Sc and vanish at Ti, along with the absence of an $N=40$ subshell at Mn. This leads to a cooler accreted neutron star crust, highlighting the connection between the structure of nuclei and neutron stars.

\end{abstract}

\maketitle

The rest mass $m$ is a basic property of an atomic nucleus, essential
for calculating astrophysical processes such as X-ray burst
light curves and $r$-process nucleosynthesis, and key to
mapping the evolution of nuclear structure across the nuclear
landscape~\citep{Scha17,Mump15,Lunn03}. While nuclear masses nearly
follow the whole-number rule, $m\approx Am_{u}$, where $A$ is the mass
number and $m_{u}=931.49$~MeV is the atomic mass unit, the $\lesssim1$\%
deviation from this relationship due to nuclear binding is notoriously difficult to predict. State-of-the-art mass
models~\citep[e.g.][]{Dufl95,Gori10,Moll12,Wang13}
often disagree in their predictions of the atomic mass excess,
${\rm ME}(Z,A)=m-(Z+N)m_{u}$ where $Z$ is the proton
number and $N$ is the neutron number, by more than one MeV.
Similar discrepancies are present when comparing predictions to experimentally measured
masses. As such, experiments mapping the evolution of the nuclear mass
surface across the nuclear landscape are essential.

For neutron-rich nuclides, mass measurements have revealed the emergence and
disappearance of the magic numbers that indicate enhanced
nuclear binding~\citep[e.g][]{Thib75,Wien13,Meis15,Rose15,Leis18,Moug18,Mich18,Xu19}. For instance,
$N=34$ semi-magicity emerges for neutron-rich calcium isotopes~\cite{Mich18}, whereas there are
signatures that the $N=40$ harmonic oscillator subshell gap disappears for neutron-rich
manganese~\cite{Naim12}. While the evolution of these subshells were mapped by spectroscopy experiments that often long-preceded the corresponding mass measurements~\citep[e.g.][]{Jans02,Lidd04,Dinc05,Craw10,Hann99,Sorl02,Gade10,Baug12}, nuclear masses provided the first model-independent confirmation of this spectroscopic evidence via fundamental ground state properties.

These evolutions in nuclear structure are closely linked to the thermal structure
of accreting neutron stars. Nuclei produced by surface burning processes are
buried by subsequent accretion, resulting in a number of nuclear reactions
in the neutron star crust that drive it from thermal equilibrium~\citep{Meis18}.
Electron-capture (EC) reactions near closed shells result in relatively large
EC-heating due to the large change in the EC $Q$-value
$Q_{\rm EC}={\rm ME}(Z,A)-{\rm ME}(Z-1,A)$~\citep{Gupt07}. EC in regions of
deformation between closed shells occur on isobars with a small odd-even staggering
in $Q_{\rm EC}$ as well as low-lying excited states, which results in the
EC-$\beta^{-}$-decay cycling process known as urca cooling~\citep{Scha14}.
Therefore, whether EC heating or cooling occurs, and the strength of the
heat source or heat sink, strongly depend on nuclear
masses~\citep{Estr11,Meis15b,Meis16}.

To simultaneously map the evolution of the $N=34$ and $N=40$ subshells and
constrain the thermal structure of accreting neutron stars, we performed
mass measurements of neutron-rich isotopes of Ar through Fe ($Z=18-26$).
First results from these measurements have been reported in
Refs.~\citep{Meis15,Meis15b,Meis16}. The present work is a re-evaluation
of the original data, incorporating recently published high-precision Penning trap mass data~\citep{Moug18,Leis18}
as additional calibration nuclides, which greatly reduces the
systematic uncertainty present in our results and greatly expands the number of nuclides for which masses are obtained. Our results extend
the known nuclear mass surface, provide model-independent confirmation of
the emergence of $N=34$ and disappearance of $N=40$ semimagicity, and
significantly update predictions for urca cooling in accreted neutron
star crusts.

Magnetic-rigidity corrected time-of-flight (TOF-$B\rho$) mass measurements
were performed at the National Superconducting Cyclotron Laboratory. The
measurement technique and measurements are described in detail in
Refs.~\cite{Mato12,Meis13,Meis15,Meis15b,Meis16} and are only briefly
summarized here. A 140~MeV/nucleon beam of $^{82}{\rm Se}$ accelerated
by the coupled cyclotrons impinged on a Be target and the resulting
fully-stripped (charge $q=Z$)
fragments were transmitted through the A1900 fragment separator~\citep{Morr03},
momentum-analyzed at the target location of the S800 spectrograph~\citep{Bazi03}, and
stopped in the focal plane of the S800~\citep{Yurk99}. Particle identification
was performed event-by-event using the TOF-$\Delta E$ method, where TOF was
provided by fast-timing scintillators separated by a flight path
$L_{\rm path}=60.6$~m and energy loss $\Delta E$ determined using an ionization
chamber. A relative measurement of $B\rho$, which is the momentum over $q$,
was obtained via a position measurement using a microchannel plate detector
located at the dispersive focus of the S800~\citep{Shap00,Roge15}.

Nominally, $m=({\rm TOF}/L_{\rm path})(qB\rho/\gamma)$, where $\gamma$
is the Lorentz factor. However, determining $L_{\rm path}$ and $B\rho$ to
sufficient precision is not practicable. Instead, an empirical relationship
between $m/q$ and TOF is determined from a fit to nuclides
of known $m$ which are simultaneously measured alongside nuclides
of interest. This work improves on prior results~\citep{Meis15,Meis15b,Meis16}
by including seven additional calibration nuclides, bringing the total to 27.
High-precision ${\rm ME}$ for $^{59-63}{\rm Cr}$~\citep{Moug18} and
$^{54,55}{\rm Ti}$~\citep{Leis18} substantially improve constraints on the
$m/q({\rm TOF})$ relationship,
whose ambiguity previously
provided one of the dominant contributions to our ME uncertainties~\citep{Meis16}.

\begin{figure}[ht!]
\begin{center}
\includegraphics[width=1.0\columnwidth]{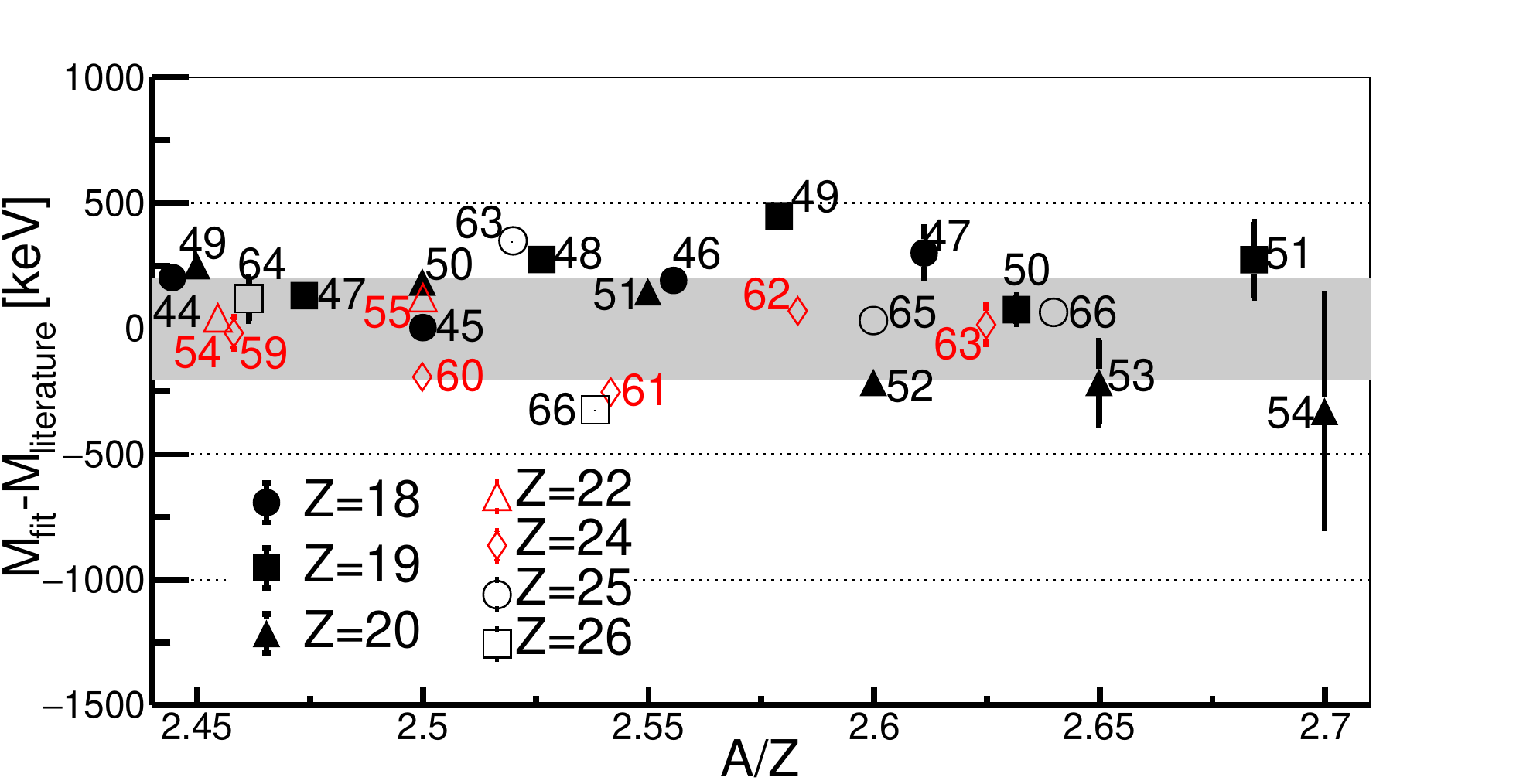}
\caption{Residuals of the $m/q(\rm TOF)$ fit to calibration nuclides, where the isotope of an element is indicated by the number next to the data point. Red data are new calibration nuclides in this re-evaluation. The gray band represents the average systematic mass uncertainty from the $\chi^{2}$ normalization.\label{fig:FitResid}}
\end{center}
\end{figure}

Mass fits were performed as described in Ref.~\cite{Meis16}. Several fit
functions were explored, of the form
\begin{equation*}
\frac{m}{q}(\tau)=a_{0}+a_{1}\tau+a_{2}z+a_{3}\tau^{2}+a_{4}z^{2}+a_{5}z\tau+f(z,\tau),
\end{equation*}
where $\tau={\rm TOF}-\langle{\rm TOF}\rangle$, $z=Z-\langle Z\rangle$, and
$f(z,\tau)$ is a function of higher-order in $z$ and/or $\tau$. The addition
of the new Ti and Cr reference nuclides resolved the previously existing
ambiguity in the $Z$-dependence, while the Cr masses additionally clarified that
a higher-order TOF component was needed to adequately minimize fit residuals,
which are shown for the best-fit in Fig.~\ref{fig:FitResid}. The best-fit
function has $f(z,\tau)=a_{6}z^{3}+a_{7}\tau^{4}$, whereas a fit of slightly
lower quality was obtained with $f(z,\tau)=a_{6}z^{3}+a_{7}\tau^{3}$.
This set of acceptable functions was determined by the following criterion.
(1) The fit residuals must lack systematic trends. (2) The fit residuals
must be robust to the arbitrary removal of reference nuclides. (3) The
difference between $\chi^{2}$ for a fit function and the best-fit
$\Delta \chi^{2}_{i}=\chi^{2}_{i}-\chi^{2}_{\rm min}$ must be within
three standard deviations of the best-fit ($\Delta \chi^{2}_{i}\lesssim37$),
which is a valid metric based on the Gaussian distribution of the fit parameters 
after repeated fits varying ME for reference nuclides in a Monte Carlo procedure~\cite{Pres92,Meis16}.

\begin{figure}[ht!]
\begin{center}
\includegraphics[width=1.0\columnwidth]{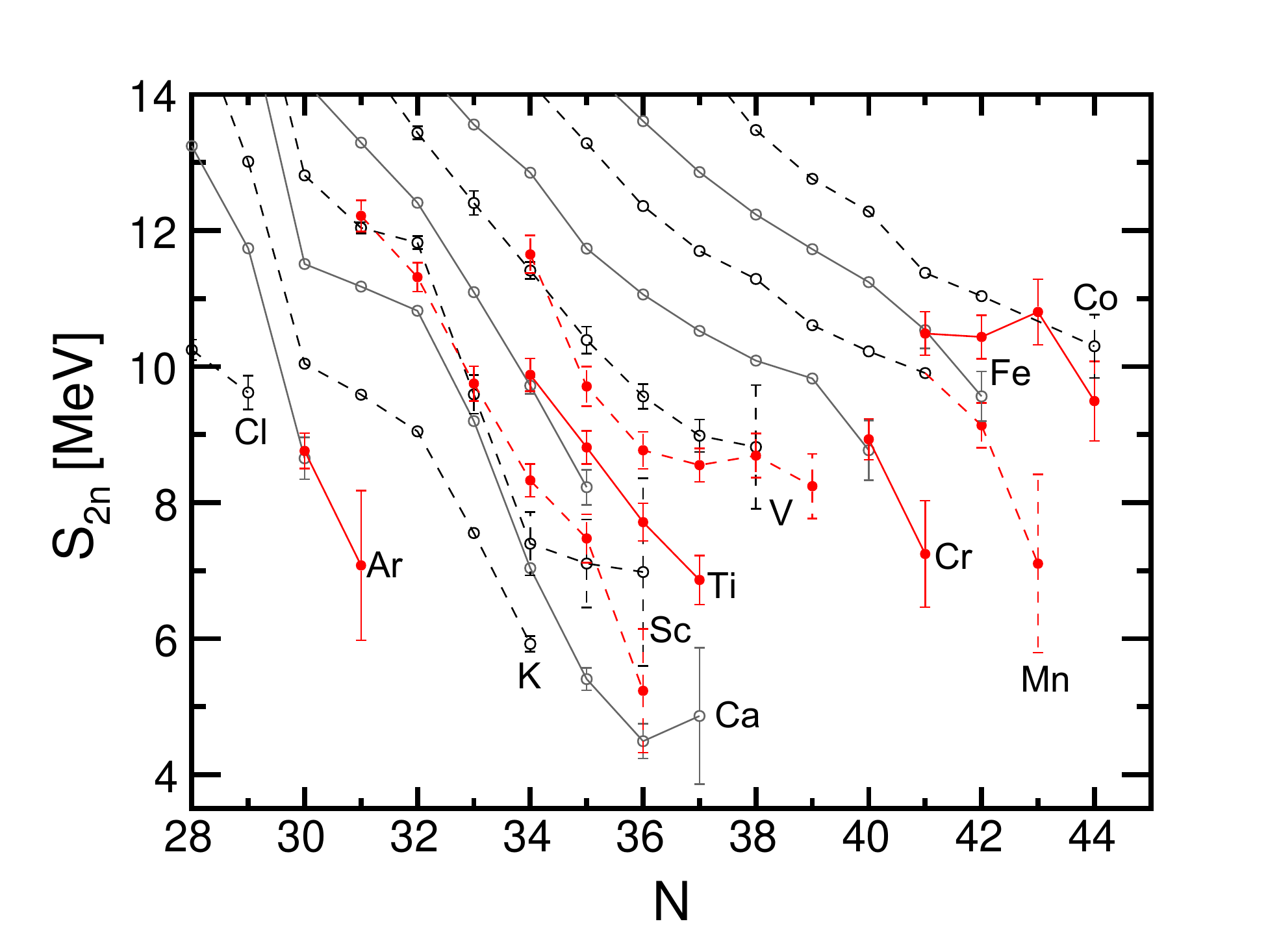}
\caption{$S_{2n}$ for isotopes of Cl (left-most trend line) through Co (right-most), where the black and gray open circles correspond to values using the 2016 AME~\citep{Wang17} or higher-precision ME from the subsequent literature~\citep{Izzo18,Moug18,Leis18,Reit18}, dashed lines are odd-$Z$, and solid lines are even-$Z$. Red filled circles are from this work.\label{fig:S2n}}
\end{center}
\end{figure}

The smaller number and closer similarity of the set of acceptable functions substantially
reduce the fit-function uncertainty relative to the previous evaluation~\citep{Meis16}.
Additional uncertainty contributions come from the 9.1~keV/$q$ systematic uncertainty
added to all nuclides to normalize the reduced $\chi^{2}$ to one for the best-fit and
the uncertainty in $a_{i}$ due to TOF uncertainties in the reference nuclides. See Ref.~\cite{Meis16} for details.

\begin{table*}[t]
  \caption{\label{tab:ME}
  Atomic mass excesses (in keV) of nuclides determined in
  this work compared to the previous evaluation~\citep{Meis15,Meis15b,Meis16}, the 2016 Atomic Mass Evaluation (AME)~\citep{Wang17}, and literature published after the 2016 AME. A * near the AME value indicates this is an extrapolation and not directly based on
  experimental data. The $I$ following an isotope indicates a known or suspected long-lived ($>100$~ns) isomeric component. For instance, for $^{67}{\rm Fe}$ the known isomer at 387~keV excitation energy is responsible for the additional asymmetric error bar, while for $^{69}{\rm Fe}$ our  results should be interpreted as an upper bound.}
  \def\arraystretch{1.25}
  \begin{ruledtabular}
  \begin{tabular}{lccccccc}
  Isotope & This Work & Previous Evaluation & AME 2016 & Literature  \\ \hline
  $^{48}{\rm Ar}$ & $-22\,390\,(260)$ & $-22\,280\,(310)$ & $-22\,280\,(310)$ & $-22\,330\,(120)$~\citep{Mich18} \\
  $^{49}{\rm Ar}$ & $-16\,300\,(1100)$ & $-17\,820\,(1100)$ & $-17\,190$*\,$(400$*$)$ & \dots\\
  $^{52}{\rm Sc}$ & $-40\,620\,(230)$ & $-40\,300\,(520)$ & $-40\,443\,(82)$ & $-40\,525\,(65)$~\citep{Xu19} \\
  $^{53}{\rm Sc}$ & $-38\,400\,(210)$ & $-38\,170\,(570)$ & $-38\,906\,(94)$ & $-38\,910\,(80)$~\citep{Xu19} \\
  $^{54}{\rm Sc}$ & $-34\,050\,(240)$ & $-33\,750\,(630)$ & $-33\,890\,(273)$ & $-34\,485\,(360)$~\citep{Xu19} \\
  $^{55}{\rm Sc}$ & $-31\,090\,(220)$ & $-30\,520\,(580)$ & $-30\,159\,(454)$ & \dots \\
  $^{56}{\rm Sc}I$ & $-25\,380\,(260)(^{+0}_{-540})$ & $-24\,850\,(590)(^{+0}_{-540})$ & $-24\,852\,(587)$ & \dots \\
  $^{57}{\rm Sc}$ & $-20\,180\,(880)$ & $-21\,010\,(1320)$ & $-20\,996\,(1304)$ & \dots \\
  $^{56}{\rm Ti}$ & $-39\,480\,(240)$ & \dots & $-39\,320\,(121)$ & $-39\,810\,(190)$~\citep{Xu19} \\
  $^{64}{\rm Cr}$ & $-33\,640\,(300)$ & $-33\,480\,(440)$ & $-33\,480\,(440)$ & \dots\\
  \end{tabular}
  \end{ruledtabular}
\end{table*}

\begin{table}[t]
  \caption{\label{tab:ME2}
  Table~\ref{tab:ME} continued, for cases without a Previous Evaluation or Literature value.}
  \def\arraystretch{1.25}
  \begin{ruledtabular}
  \begin{tabular}{lccccc}
  Isotope & This Work  & AME 2016  \\ \hline
  $^{57}{\rm Ti}$ & $-34\,500\,(240)$ & $-33\,916\,(256)$ \\
  $^{58}{\rm Ti}$ & $-30\,890\,(250)$ & $-31\,110$*$\,(200$*$)$ \\
  $^{59}{\rm Ti}$ & $-25\,220\,(270)$ & $-25\,510$*\,$(200$*$)$ \\
  $^{57}{\rm V}$ & $-44\,650\,(260)$ & $-44\,413\,(80)$ \\
  $^{58}{\rm V}$ & $-39\,720\,(230)$ & $-40\,402\,(89)$ \\
  $^{59}{\rm V}$ & $-37\,040\,(260)$ & $-37\,832\,(162)$ \\
  $^{60}{\rm V}I$ & $-32\,810\,(230)(^{+0}_{-202})$ & $-33\,242\,(220)$ \\
  $^{61}{\rm V}$ & $-30\,380\,(280)$ & $-30\,506\,(894)$ \\
  $^{62}{\rm V}$ & $-25\,340\,(420)$ & $-25\,476$*\,$(298$*$)$ \\
  $^{65}{\rm Cr}$ & $-27\,280\,(780)$ & $-28\,220$*\,$(300$*$)$ \\
  $^{67}{\rm Mn}$ & $-33\,960\,(330)$ & $-33\,460$*\,$(300$*$)$ \\
  $^{68}{\rm Mn}$ & $-27\,710\,(1310)$ & $-28\,380$*\,$(400$*$)$ \\
  $^{67}{\rm Fe}I$ & $-45\,560\,(320)(^{+0}_{-387})$ & $-45\,610\,(270)$ \\
  $^{68}{\rm Fe}$ & $-44\,360\,(320)$ & $-43\,487\,(365)$ \\
  $^{69}{\rm Fe}I$ & $-40\,270\,(400)(^{+0}_{-?})$ & $-39\,030$*\,$(400$*$)$ \\
  $^{70}{\rm Fe}$ & $-37\,710\,(490)$ & $-36\,510$*\,$(400$*$)$ \\
  \end{tabular}
  \end{ruledtabular}
\end{table}

Our resultant ME are reported in Tabs.~\ref{tab:ME} and \ref{tab:ME2} with comparisons to literature values and results
from our previous evaluation, where Fig.~\ref{fig:S2n} shows the two neutron
separation energy $S_{2n}(Z,A)=2{\rm ME}(0,1)+{\rm ME}(Z,A-2)-{\rm ME}(Z,A)$. All but one of our updated ME are within 1 standard deviation $\sigma$ of
our previous ME and all are within 2$\sigma$, while the majority of uncertainties have been reduced
by a factor of two. We report ME for $^{58,59}{\rm Ti}$, $^{62}{\rm V}$, $^{65}{\rm Cr}$,
$^{67,68}{\rm Mn}$, and $^{69,70}{\rm Fe}$ for the first time. 

The new trend in $S_{2n}$ for V is largely due to $^{58,59}{\rm V}$, which are much less bound in our work compared to the 2016 AME evaluated result, but in agreement with the privately communicated results Ref.~\cite{Wang17} refers to as {\tt 1998Ba.A} that were included in that evaluation. The abnormal behavior for
Fe in $S_{2n}$ shown in Fig.~\ref{fig:S2n} is difficult to understand in terms of nuclear
structure effects, since a new single-particle orbital is not expected to be filled. Additionally,
anomalous behavior of $m/q({\rm TOF})$ is an unlikely explanation,
since the function is smooth in that region and a similar feature is not seen for nearby $Z$. $^{67}{\rm Fe}$ is known to have an isomeric state at 387~keV~\cite{Sawi03}, which, along with measurement uncertainties, may explain the $\sim1$~MeV deviation in $S_{2n}$ from a smooth trend. We suspect $^{69}{\rm Fe}$ may also have a long-lived isomer based on the presence of such
states in odd-$A$ isotopes of Fe due to intrusion from the $\nu g_{9/2}$ orbital~\cite{Luna07}. The kink at $N=39$ for Cr agrees with the trend calculated using the LNPS' Hamiltonian~\citep{Lenz10,Meis16,Moug18}, but the absolute $S_{2n}$ are discrepant.

\begin{figure}[ht!]
    \centering
     \subfigure{\label{fig:Dn34}\includegraphics[width=0.5\textwidth]{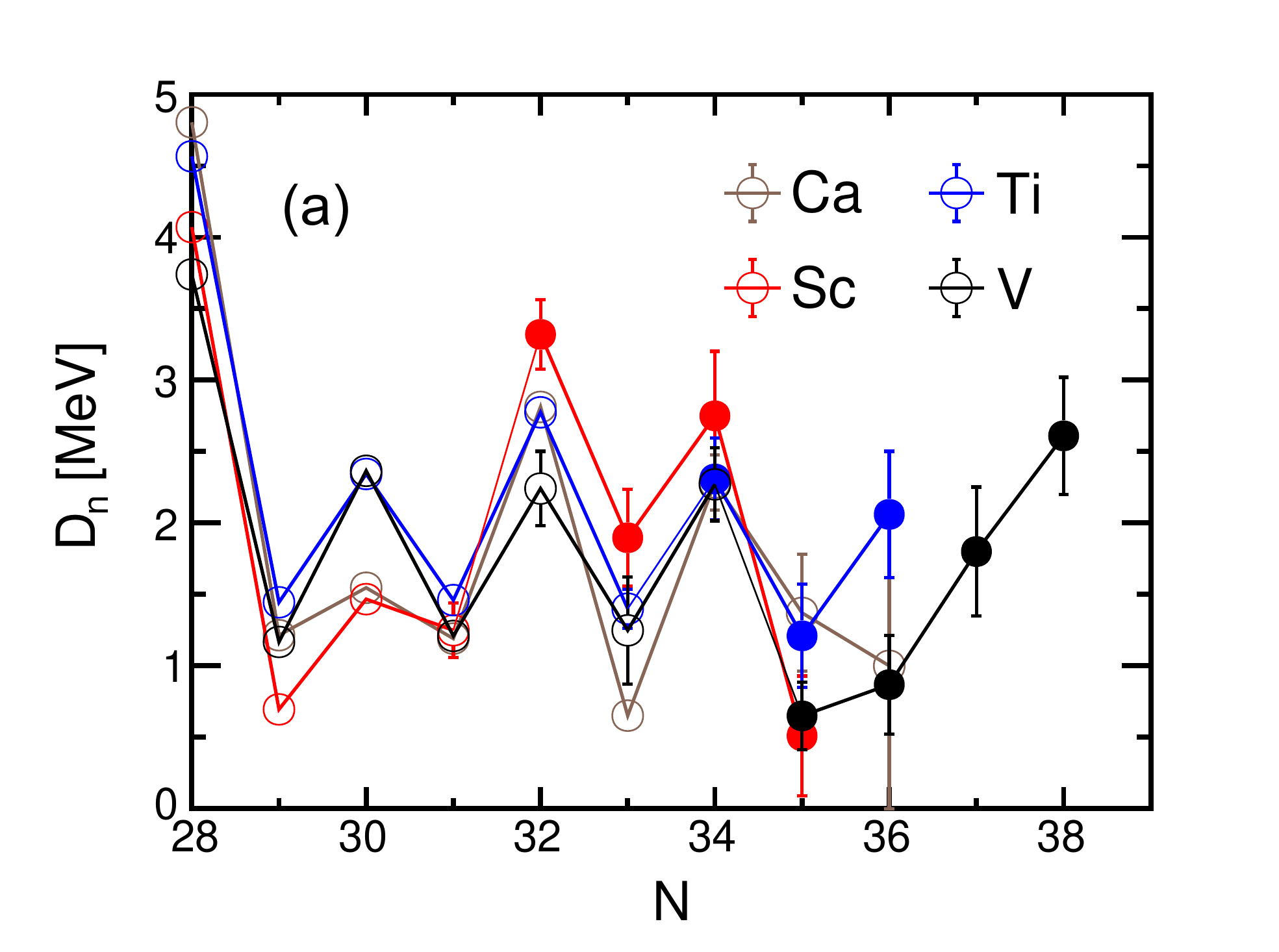}}
  \qquad
     \subfigure{\label{fig:Dn34shell}\includegraphics[width=0.5\textwidth]{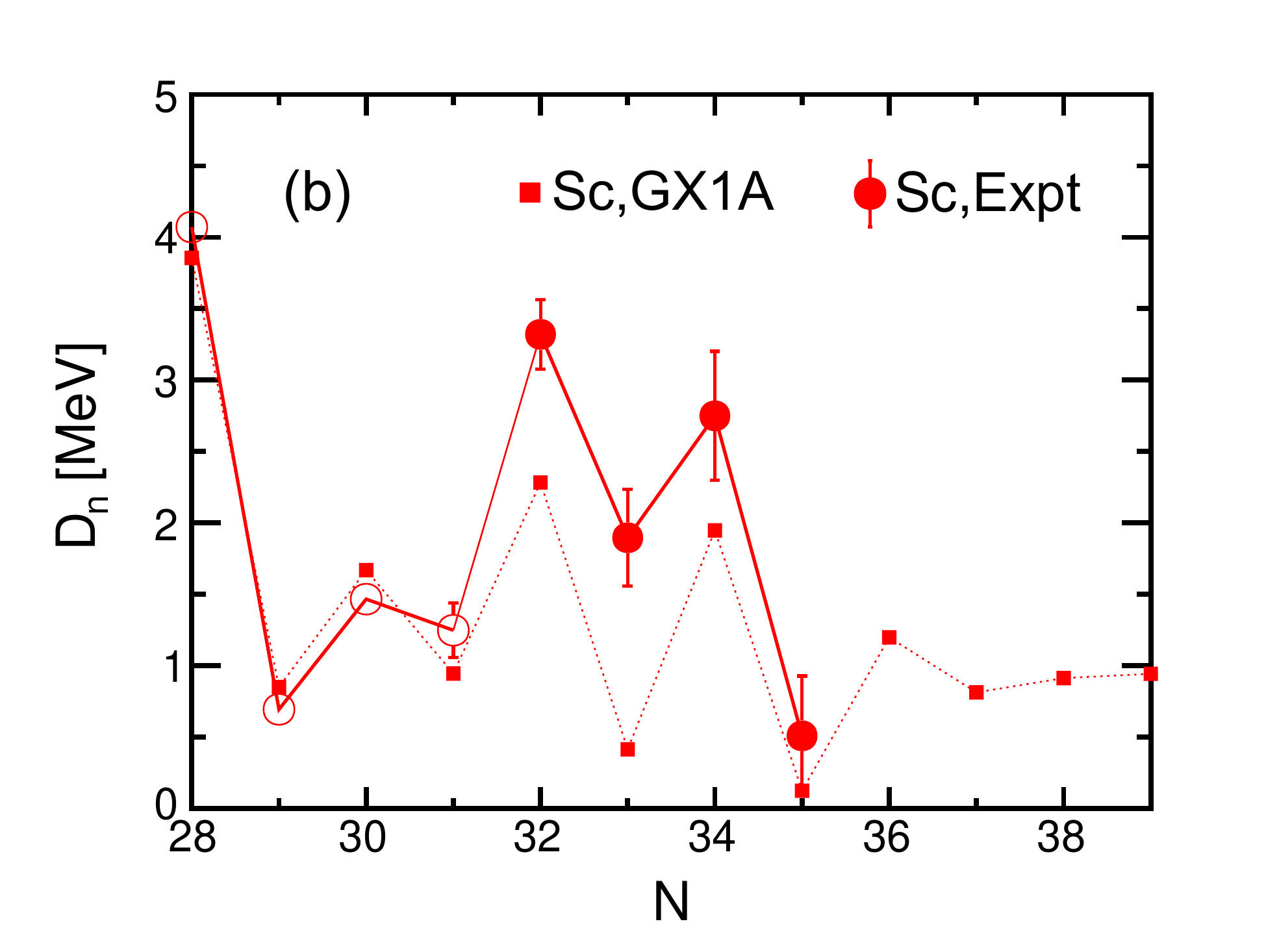}}
  \qquad
  \subfigure{\label{fig:Dn40}\includegraphics[width=0.5\textwidth]{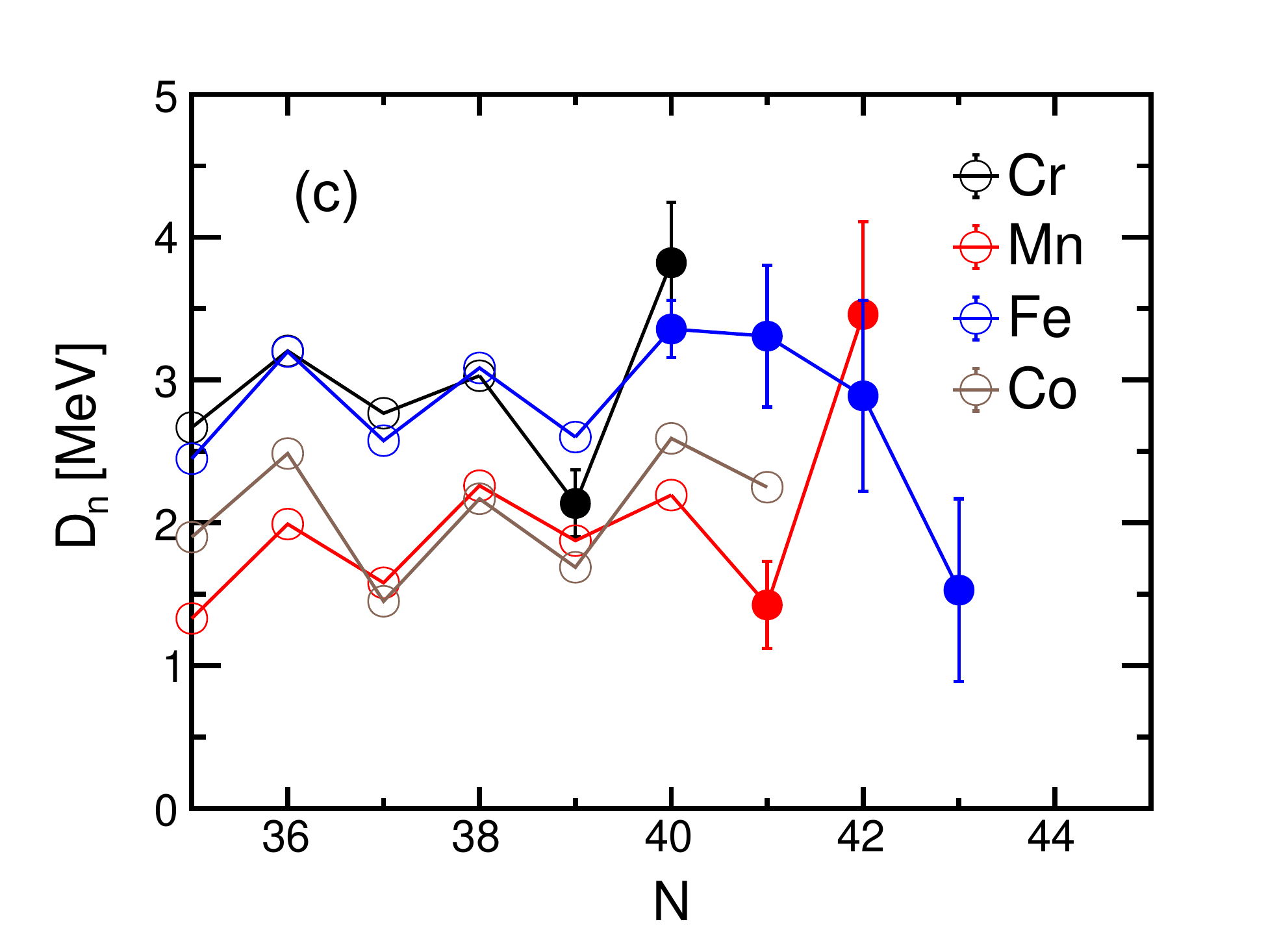}}
    \caption{$D_{n}$ 
    using our ME (filled circles) and ME from the literature (open circles) (a) near $N=34$ and (c) near $N=40$. A comparison to shell model calculations with the GX1A Hamiltonian (filled-squares) is shown for Sc in (b).}
    \label{fig:Dn}
\end{figure} 

ME for $N=36$ isotopes of Sc, Ti, and V can be used to deduce the evolution of $N=34$ in this region.
Ref.~\citep{Mich18} demonstrated the presence of $N=34$ semi-magicity for Ca, whereas spectroscopy has indicated this subshell closure is absent for Ti and likely weakened for Sc, with the caveat that $E(2_{1}^{+})$ energies can provide ambiguous constraints on shell gaps~\cite{Step17}. Our data reveal a continuous slope in $S_{2n}$ for Ti following $N=34$, while Sc trends slightly more negative beyond this point. This indicates that a weak $N=34$ subshell gap is present at Sc. 

Further insight is provided
by the trend in $D_{n}(Z,A)=(-1)^{N+1}[S_{n}(Z,A+1)-S_{n}(Z,A)]$, which is related to the empirical pairing gap~\citep{Brow13}. $D_{n}$ is proportional to the number of angular momentum projection states ($2j+1$) participating in pairing, providing a signature of gaps in single particle levels. Fig.~\ref{fig:Dn}(a) shows the trend in $D_{n}$ near $N=34$ for isotopes of Ca through V. While the dip after $N=34$ for the Sc isotopes might initially appear to be the signature of a significant shell gap, shell model calculations suggest this is not the case. Calculations using the GX1A Hamiltonian~\cite{Honm05}, whose results are shown in Fig.~\ref{fig:Dn}(b) to be in qualitative agreement with experiment, indicate that the particularly low $D_{n}$ for $^{56}{\rm Sc}$ is due to the large splitting of levels created by the residual interaction between the $\pi f_{7/2}$ and $\nu f_{5/2}$ orbitals. In particular, the minimum in $D_{n}$ for Sc is due to the low-lying $J=1$ level created by this interaction. $S_{2n}$ is not sensitive to this effect as it reflects the energetics of neutron shells and not the strength of proton-neutron pairing~\cite{Lunn03,Naim12}. These results confirm previous indications from spectroscopy of weak $N=34$ magicity for Sc, while removing the ambiguity inherent to spectroscopic interpretations of shell structure~\cite{Lidd04,Craw10,Step17}.

Fig.~\ref{fig:S2n} demonstrates a continuous slope in $S_{2n}$ through $N=40$ for Mn, strengthening
the conclusions of Ref.~\citep{Naim12} that the $N=40$ subshell is absent for this element. This is bolstered by the trend in $D_{n}$ shown in Fig.~\ref{fig:Dn}(c). Our data are ambiguous regarding the $N=40$ subshell at Cr, where the mass of $^{66}{\rm Cr}$ is needed to confirm prior spectroscopic and coulomb-excitation evidence~\citep[e.g.][]{Sorl03,Gade10,Craw13}.

The evolution in nuclear structure presented here is directly linked to the thermal structure of
accreting neutron stars. The neutrino luminosity from urca cooling $L_{\nu}\propto X(A)(ft)^{-1}|Q_{\rm EC}|^{5}$,
where $ft$ is the comparative half-life and $X(A)$ is the mass fraction,
and is therefore very sensitive to changes in ME~\citep{Tsur70,Deib16}. This process operates 
in the accreted neutron star crust with consequential $L_{\nu}$ for odd-$A$ nuclides with
$X(A)\gtrsim0.5$\%, $ft\lesssim5$, and $8\lesssim |Q_{\rm EC}|\lesssim15$~MeV~\citep{Scha14,Meis17},
where the upper limit on $Q_{\rm EC}$
is due to competing EC reaction channels~\citep{Gupt08}. Two EC parents predicted to produce
some of the largest $L_{\nu}$, with potentially observable consequences~\citep{Deib16,Meis17},
are $^{55}{\rm Sc}$ and $^{65}{\rm Mn}$. For $^{65}{\rm Mn}$, $ft$ will be
uncertain by orders of magnitude until measurements are enabled by next-generation rare isotope beam facilities,
but estimates from QRPA calculations~\citep{Scha14} and using the Moszkowski nomographs~\citep{Mosz51} result
in $\log(ft)\approx5$, allowing significant $L_{\nu}$. $^{55}{\rm Sc}$ by contrast has more
consistent predictions, with $\log(ft)\approx5$ using QRPA methods~\citep{Scha14}, the Moszkowski nomographs, and
empirical systematics~\citep{Sing98}, currently making it the highest predicted $L_{\nu}$ urca cooling layer
~\citep{Meis17,Meis18c}. Furthermore, a measurement of $ft$ for this transition has recently
taken place~\citep{Ong19}. For both $A=55$ and 65, $X(A)>0.5$\% and are remarkably consistent for a wide variety
of assumptions for nuclear burning on the accreting neutron star surface~\citep{Meis19}. Therefore $Q_{\rm EC}$ are
the final piece of the nuclear physics puzzle for these urca coolers.

Using the newly determined ${\rm ME}(^{55}{\rm Sc})$, $|Q_{\rm EC}(^{55}{\rm Sc})|=12.44\,(0.27)$~MeV, to be compared
to the prior~\citep{Wang17,Mich18} value  11.51\,(0.48)~MeV. Our reported ${\rm ME}(^{65}{\rm Cr})$ results in
$|Q_{\rm EC}(^{65}{\rm Mn})|=13.69\,(0.78)$~MeV, to be compared to the prior~\citep{Naim12,Wang17} value
12.75\,(0.30)~MeV (though the latter uncertainty relies on the AME extrapolation, which
assumes an essentially featureless nuclear mass surface and may be underestimated).
These results increase $L_{\nu}$
by 50\% for EC on $^{55}{\rm Sc}$ with half the prior uncertainty and provide the first experimental determination of $L_{\nu}$
for EC on $^{65}{\rm Mn}$. While in agreement with previous predictions, our central value for $Q_{\rm EC}(^{65}{\rm Mn})$ leads to 40\% larger $L_{\nu}$. Therefore the accreted neutron star
crust is cooler than previously thought, with an improved precision on the description of the neutron star
thermal structure. 

In conclusion, this work highlights the intriguing connection between evolution in nuclear structure and the thermal structure
of accreting neutron stars. We find model-independent evidence for the onset of the $N=34$ subshell for Sc and the likely absence of $N=40$ magicity for Cr, each of which result in a larger mass difference for transitioning from
odd-$Z$ to odd-$N$ in EC. This
is ultimately connected
to the strength of the interaction between a nuclear core and an unpaired proton as opposed to an unpaired neutron~\citep{Zeld65,Naim12},  and 
leads to increasing the phase-space available for the weak transitions involved
in urca cooling, which in turn results in a cooler neutron star crust. Our measurements leave ${\rm ME}(^{63}{\rm V})$ as the final nuclear mass important for urca cooling
in the accreted neutron star crust that relies on theoretical mass estimates.

\begin{acknowledgments}
This work was funded by the U.S. Department of Energy
Office of Science through Grants No. DE-FG02-88ER40387, DE-SC0019042, DE-SC0020451, and DE-SC0020406; the U.S. National Science Foundation through Grants No. PHY-0822648,
PHY-1102511, PHY-1811855, PHY-1913554, and PHY-1430152 (Joint Institute for Nuclear Astrophysics -- Center for the Evolution of the Elements); and the DFG under Contracts No. GE2183/1-1 and No. GE2183/2-1.
\end{acknowledgments}

\bibliographystyle{apsrev4-2}
\bibliography{References}

\end{document}